\documentclass[twocolumn,showpacs,amsmath,amssymb,prl]{revtex4}

\usepackage{graphicx}
\usepackage{dcolumn}
\usepackage{bm}
\usepackage{psfrag}

\usepackage{latexsym}

\usepackage{amsmath,amssymb,dsfont,mathrsfs}

\newcommand{\psibar}{\bar{\psi}}

\def\slashchar#1{\setbox0=\hbox{$#1$} 
\dimen0=\wd0 
\setbox1=\hbox{/} \dimen1=\wd1 
\ifdim\dimen0>\dimen1 
\rlap{\hbox to \dimen0{\hfil/\hfil}} 
#1 
\else 
\rlap{\hbox to \dimen1{\hfil$#1$\hfil}} 
/ 
\fi}

\begin{document}

\preprint{}

\title{
A Z$_2$ 
index of Dirac operator with time reversal symmetry
}

\author{Takahiro Fukui and Takanori Fujiwara}
 \affiliation{Department of Physics, Ibaraki University, Mito
 310-8512, Japan}

\date{\today}

\begin{abstract}
With time reversal symmetry a Dirac operator has vanishing index and Chern number.
We show that
we can nevertheless define a nontrivial Z$_2$ index as well as a corresponding 
topological invariant given by gauge field, 
which implies that such a Dirac operator is topologically nontrivial.
\end{abstract}


\maketitle

The idea of topological invariants has been successfully applied 
to various fields in physics. In gauge theories they 
are used to classify topological configurations such as 
monopoles and instantons. When couplings with chiral 
fermions are introduced, there arise interesting field theoretical 
phenomena such as chiral anomaly and gauge anomaly \cite{AnoRev}. 
The chiral anomaly \cite{AnoRev} is known to have intimate relationship with 
the index theorem \cite{ABP73} which tells that the index 
of the Dirac operator coincide with the second Chern number.
The gauge anomaly also has topological origin, since it is 
related with the chiral anomaly in six dimensions \cite{AlvGin84}. 

In condensed matter physics, 
it is well-known that the plateaus of the quantum Hall effect (QHE) are classified 
by the first Chern number \cite{TKNN82,Koh85}. Recently,  
a novel topological number has been proposed by Kane, Mele, and Fu
 \cite{KanMel05a,KanMel05b,FuKan06}
for quantum spin Hall effect (QSHE) \cite{KanMel05a,KanMel05b,FuKan06,BerZha06}.
It is invariant only modulo two, and is often called Z$_2$ invariant.  
Here, time reversal symmetry is the key difference between QHE and QSHE.
Remarkably, QSHE have recently been observed in several experiments
\cite{KWBetal07,HsiQia08,HsiXia09,NisTas09}.

The formula of the Z$_2$ invariant proposed 
by Fu and Kane \cite{FuKan06} is, roughly speaking, ``half'' 
the first Chern number. Therefore, 
it is very useful \cite{FukHat07} for numerical calculations if we utilize the techniques 
of computing the Chern number in the lattice gauge theories \cite{Luscher99}. 
Besides such practical applications, it is of fundamental importance, since 
it could be a topological invariant in  ``mod 2 index theorem'' \cite{AtiSin71}.
Therefore, if we find a corresponding analytical invariant, 
we can obtain a simple formula of a Z$_2$ index theorem for a (pseudo-real) Dirac operator. 

In this paper, we investigate analytical and  topological invariants associated 
with a Dirac operator with time reversal invariance.
We first study its spectral properties in Euclidean space and
define a Z$_2$ index of the Dirac operator. 
We then propose a topological invariant which is a generalization 
of the Fu-Kane formula, and infer that it coincides with the Z$_2$ index.

We begin by recalling the time reversal transformation of Dirac fermions in 
$d=2n+2$ ($n=0,1,\cdots$) dimensional Minkowski space-time. 
It is defined by $\psi(t,\bm x)\rightarrow {\cal T}\psi(-t,\bm x)$, where 
${\cal T}$ is an anti-unitary operator, 
${\cal T}\equiv \Gamma_\gamma\Gamma_{\rm G}{\cal K}$,
with $\Gamma_\gamma$ being a product of some $\gamma$ matrices, 
$\Gamma_{\rm G}$ a generator of a gauge group G, 
and ${\cal K}$ the operator of taking complex conjugate. 
For the Lagrangian density ${\cal L}(t,\bm x)=\psibar(t,\bm x)i\slashchar{D}(t,\bm x)\psi(t,\bm x)$
to transform as 
${\cal L}(t,\bm x)\rightarrow {\cal L}(-t,\bm x)$ under time reversal, we see
${\cal T}i\slashchar{D}(t,\bm x){\cal T}^{-1}=i\slashchar{D}(-t,\bm x)$, from which
it follows that
\begin{alignat}1
&
{\cal T}\gamma^\mu{\cal T}^{-1}=\gamma_\mu,
\nonumber\\
&
{\cal T}A_\mu(t,\bm x){\cal T}^{-1}=A^\mu(-t,\bm x),
\label{TimRevMin}
\end{alignat}
where the metric is $g_{\mu\nu}={\rm diag}(1,-1,\cdots,-1)$.
The $\Gamma_5$ matrix anti-commuting with the Dirac operator is given by 
$\Gamma_5=i^{d/2-1}\gamma^0\gamma^1\cdots\gamma^{d-1}$.
This definition directly leads to 
${\cal T}\Gamma_5{\cal T}^{-1}=-\Gamma_5$ for $d=4n+2$ and
$=+\Gamma_5$ for $d=4n+4$.

${\cal T}$ has the following two possibilities: 
${\cal T}^2=\pm1$, depending on $\Gamma_\gamma^2=\pm1$ and $\Gamma_{\rm G}^2=\pm1$.
The former is determined solely by the space-time dimension $d$: 
In $d=2$, for example, we can choose $\gamma^0=\sigma^2$ and $\gamma^1=i\sigma^1$.
Since these are imaginary, ${\cal K}(\gamma^0,\gamma^1){\cal K}^{-1}=(-\gamma^0,-\gamma^1)$, 
we see that $\Gamma_\gamma=-i\gamma^1=\sigma^1$, and therefore, $\Gamma_\gamma^2=1$.
In general, we have
$\Gamma_\gamma^2=1$ for $d=0, 2+8n$, and $\Gamma_\gamma^2=-1$ for $d=4, 6+8n$.

To discuss the index of the Dirac operator with time reversal invariance, we switch 
from Minkowski space to Euclidean space.
It should be noted that a Euclidean version of the time reversal transformation is not so obvious,
since it includes the operator ${\cal K}$. 
Here, we define Euclidean space by rotating all the spatial coordinates
$x^j$ ($j=1,\cdots,d-1$) onto the imaginary axes via $x^j=iy^j$, whereas $x^0=y^d$.
As we shall see, this enables us to relate a Z$_2$ index of the Dirac operator with
a topological invariant \cite{footnote1}.
The metric becomes in this case $g_{\mu\nu}=\delta_{\mu\nu}$.
Correspondingly, the $\tilde\gamma$ matrices are introduced via 
$\gamma^j=i\tilde\gamma^j$ 
and $\gamma^0=\tilde\gamma^d$ which become hermitian 
$\tilde\gamma^{\mu\dagger}=\tilde\gamma^\mu$, and the gauge potential ${\cal A}_\mu(y)$
via $A_j(x)=-i{\cal A}_j(y)$ and $A_0(x)={\cal A}_d(y)$. Then, the Dirac operator
can be denoted as 
$i\slashchar{D}(x)=i\slashchar{D}(y)\equiv i\tilde\gamma^\mu(\partial_{y^\mu}-i{\cal A}(y))$
which we regard as hermitian $(i\slashchar{D}(y))^\dagger=i\slashchar{D}(y)$.
The transformation law under time reversal becomes
\begin{alignat}1
&
{\cal T}\tilde\gamma^\mu{\cal T}^{-1}=\tilde\gamma^\mu,
\nonumber\\
&
{\cal T}{\cal A}_\mu(y){\cal T}^{-1}={\cal A}_\mu(-y),
\label{TimRevInvEuc}
\end{alignat}
which follows from the same transformation law (\ref{TimRevMin}) 
but with the Euclidean metric mentioned above. 
Therefore, the Dirac operator transforms as
\begin{alignat}1
{\cal T}i\slashchar{D}(y){\cal T}^{-1}=i\slashchar{D}(-y).
\label{DirUndT}
\end{alignat}
Note that this transformation is for flat space: If we consider curved space,
it should be modified suitably, as we shall see.
Other key symmetry is chiral symmetry described by
\begin{alignat}1
\Gamma_5 i\slashchar{D}(y)+i\slashchar{D}(y)\Gamma_5=0 .
\label{ChiSym}
\end{alignat}

The Z$_2$ index discussed in this paper is involved in the case
with both of the conditions
\begin{subequations}
\label{allequation}
\begin{alignat}1
&
{\cal T}^2=-1, 
\label{TSqu}
\\
&
{\cal T}\Gamma_5{\cal T}^{-1}=-\Gamma_5 
\label{TGam5}
\end{alignat}
\end{subequations}
fulfilled.
The former ensures that the eigenstates are always doubly-degenerate, which 
is referred to as Kramers doublet. The latter claims that
the zero-mode Kramers doublet have opposite chiralities.
These conditions give some constraints:
Eq. (\ref{TGam5}) is valid only in $d=4n+2$, and 
Eq. (\ref{TSqu}) imposes $\Gamma_{\rm G}^2=-1$ for $d=8n+2$ and
$\Gamma_{\rm G}^2=+1$ for $d=8n+6$.  
Typical example in the former case is $\Gamma_{\rm G}=1_N\otimes i\tau^2\equiv J_2$, 
whereas in the latter case, a convenient but nontrivial choice may be 
$\Gamma_{\rm G}=1_N\otimes\tau^1\equiv J_1$ \cite{footnote2}. 
The transformation law of gauge potentials is thus defined by
\begin{alignat}1
{\cal A}_\mu(-y)=\left\{
\begin{array}{l}
J_2{\cal A}_\mu^*(y)J_2^{-1}\\ J_1{\cal A}_\mu^*(y)J_1^{-1}
\end{array}
\right.
\,\,{\rm for}\,\,d=8n+
\left\{
\begin{array}{l}
2\\6
\end{array}
\right. .
\label{TimRevInvGau}
\end{alignat}

For a time reversal invariant Dirac operator discussed so far, 
we shall define a Z$_2$ index.
Let $\varphi_k(y)$ be an eigenstates of $i\slashchar{D}$:
\begin{alignat}1
i\slashchar{D}(y)\varphi_k(y)=\varepsilon_k\varphi_k(y) .
\nonumber
\end{alignat}
Then, Eq. (\ref{DirUndT}) ensures that $\varphi_{{\rm K}k}(y)\equiv{\cal T}\varphi_k(-y)$ 
is also an eigenstate of $i\slashchar{D}(y)$ with the same eigenvalue $\varepsilon_k$.
\begin{figure}[h]
\begin{center}
\includegraphics[width=0.5\linewidth]{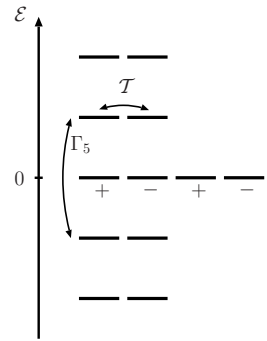}
\caption{
Schematic illustration of the spectrum of the Dirac operator in the case $N_{\rm K}=2$
(two zero-mode doublets) on a compact manifold M. 
The nonzero-mode quartet can be obtained by the operation
of ${\cal T}$ and/or $\Gamma_5$. 
The chirality $\pm$ is shown for the zero-mode eigenstates.
}
\label{f:Spe}
\end{center}
\end{figure}
Here, the condition ({\ref{TSqu}) plays a vital role in the orthogonality 
between $\varphi_k$ and $\varphi_{{\rm K}k}$. It thus turns out that all eigenstates
are doubly-degenerate, called Kramers doublets as mentioned above, 
which we denote as $\Phi_k(y)=(\varphi_k(y),\varphi_{{\rm K}k}(y))$.
The spectrum is illustrated in Fig. \ref{f:Spe}.

Let us concentrate on the zero-mode eigenstates, $\Phi_{0,\alpha}$ ($\alpha=1,\cdots,N_{\rm K}$). 
Since the Dirac operator anti-commutes with $\Gamma_5$, the zero-modes can be 
chosen to be eigenstates of the chirality. Suppose
$\Gamma_5\varphi_{0,\alpha}=+\varphi_{0,\alpha}$. Then, we see
$\Gamma_5\varphi_{{\rm K}0,\alpha}=-\varphi_{{\rm K}0,\alpha}$ because of Eq. (\ref{TGam5}).
Namely, at the zero energy, each Kramers doublet is composed of two states with
opposite chiralities. 
Even when there are some doublets at the zero-energy, the number of 
states with positive chirality is the same as the number of states with negative chirality:
\begin{alignat}1
{\rm ind}\, i\slashchar{D}&\equiv 
{\rm dim}\,{\rm ker}\,i\slashchar{D}_+-{\rm dim}\,{\rm ker}\,i\slashchar{D}_-=0 ,
\nonumber
\end{alignat}
where $i\slashchar{D}_\pm\equiv i\slashchar{D}P_\pm$ with $P_\pm\equiv (1\pm\Gamma_5)/2$.
The index of the present Dirac operator is thus trivial. 
Nevertheless, the time reversal invariance (\ref{DirUndT}), 
if combined with chiral symmetry (\ref{ChiSym}), gives an interesting invariant. 
Chiral symmetry (\ref{ChiSym}) tells that if $\Phi_k$ is an 
eigen-doublet with the energy $\varepsilon_k$, the state defined by
$\Phi_{-k}\equiv\Gamma_5\Phi_k$ is also an eigen-doublet with the opposite energy $-\varepsilon_k$.
Therefore, nonzero-mode states form a quartet in this sense.
Suppose that we have just one Kramers doublet at the zero energy. Then, it turns out 
that this doublet is stable against perturbations with time reversal and chiral symmetries,
since these two states cannot move to nonzero energies without two more states  
in order to ensure both the symmetries.
On the other hand, if there are two doubles at the zero-energy, they are not obliged to 
stay there: Small perturbations enable two of them to move to the positive energies 
and the other two to move to the opposite negative energies.
In more general, we can claim that {\it evenness} or {\it oddness} of the number of the 
zero-mode Kramers doublets is an analytic invariant, from which 
we define a Z$_2$ index of the Dirac operator with time reversal symmetry,
\begin{alignat}1
{\rm ind}_+\, i\slashchar{D}&\equiv {\rm dim}\,{\rm ker}\,i\slashchar{D}_+ \quad{\rm mod}\,\,2 .
\label{Z2Ind}
\end{alignat}

Next, we define a topological invariant given by the gauge field.
To this end, we must first specify the manifold compatible with the condition 
(\ref{DirUndT}). Let M be a compact manifold without boundary.
We assume that it can be divided into two M$_\pm$ such that 
if $y\in$ M$_+$, $-y\in$ M$_-$ except for the time reversal invariant points
$y_j=-y_j$ ($j=1,\cdots,N_{\rm inv}$). The number $N_{\rm inv}$ of such points depends on M. 
\begin{figure}[h]
\begin{center}
\includegraphics[width=0.5\linewidth]{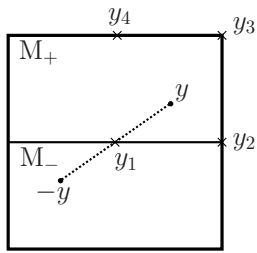}
\caption{
An example of two-dimensional manifold M.
$y_j$ denotes the time reversal invariant points.
The square represents ${\rm S}^2$ if the boundary 
is regarded as one point. In this case, time reversal invariant points are just two, 
$y_1$ and $y_2(=y_3=y_4)$. If the two parallel boundaries 
are pasted and the periodic boundary conditions is imposed on each direction, 
the same square now denotes ${\rm T}^2$, which has four 
time reversal invariant points. 
}
\label{f:M}
\end{center}
\end{figure}
For example, the two-sphere ${\rm S}^2$ has two invariant points, whereas the 
two-torus ${\rm T}^2$ has four invariant points, as illustrated in Fig. \ref{f:M}.

To define a topological invariant, it is convenient to
define a gauge potential one-form ${\cal A}=-i{\cal A}_\mu {\rm d}y^\mu$ 
and corresponding field strength two-form
${\cal F}={\rm d}{\cal A}+{\cal A}^2$.
Time reversal invariance (\ref{TimRevInvEuc}) or (\ref{TimRevInvGau}) tells that
the Chern number $C_{d}$ ($d=4n+2$ with $n=0,1,\cdots$) is vanishing:
\begin{alignat}1
C_{d}={\mathscr N}_{d/2}\int_{{\rm M}}{\rm tr}\,{\cal F}^{d/2}=0 ,
\nonumber
\end{alignat}
where the numerical factor is ${\mathscr N}_{n}\equiv i^{n}/[n!(2\pi)^{n}]$.
This is consistent with the spectral property of the Dirac operator whose index is zero,
as discussed above. This feature is quite similar to the QSHE:
The first Chern number associated with the Berry phase in the Brillouin zone (${\rm T}^2$)
vanishes due to time reversal symmetry. Nevertheless, the QSHE phase is the topologically
nontrivial phase which can be described by the Z$_2$ number,
as shown by Kane {\it et. al.} \cite{KanMel05b,FuKan06}.
Motivated by their work, we propose that 
the Z$_2$ index (\ref{Z2Ind}) in $d=4n+2$ is equivalent to
\begin{alignat}1
D_d={\mathscr N}_{d/2}
\left[
\int_{{\rm M}_+}{\rm tr}\,{\cal F}^{d/2}
-\int_{\partial{\rm M}_+}\omega_{d-1}({\cal A})
\right] ,
\label{Z2IndThe}
\end{alignat}
mod 2. Here, $\omega_{d-1}$ is the Chern-Simons $(d-1)$-form
which obeys ${\rm tr}\,{\cal F}^{d/2}={\rm d}\omega_{d-1}({\cal A})$ \cite{Zum}.
It should be noted that the formula (\ref{Z2IndThe}) have meaning under 
the condition (\ref{TimRevInvGau}). 
This is the reason we have adopted an unconventional Wick rotation \cite{footnote1}.

We must first examine the gauge-dependence of $D_d$.
Let ${\cal A}_g=g^{-1}{\cal A}g+g^{-1}{\rm d}g$ be the gauge-transform of ${\cal A}$.
If the time reversal invariance (\ref{TimRevInvGau}) is enforced on ${\cal A}_g$,
$g$ should obey
$g(-y)=J_2g^*(y)J_2^{-1}$ for $d=8n+2$ or $g(-y)=J_1g^*(y)J_1^{-1}$ for $d=8n+6$.
We will refer to this condition as the time reversal constraint on the gauge transformation.
Let $\Delta_{d-1}[g]\equiv D_{d}[{\cal A}_g]-D_{d}[{\cal A}]$ be the gauge-dependence of $D_d$.
Note 
\begin{alignat}1
\omega_{d-1}({\cal A}_g)-&\omega_{d-1}({\cal A})=\omega_{d-1}(g^{-1}{\rm d}g)+{\rm d}\alpha_{d-2},
\nonumber
\end{alignat}
where $\alpha_{d-2}$ is a $(d-2)$-form \cite{Zum}, which leads to 
\begin{alignat}1
\Delta_{d-1}[g]=\frac{(-1)^{n+1}i}{(2\pi)^{2n+1}}\frac{(2n)!}{(4n+1)!}\int_{\partial{\rm M}_+}
{\rm tr}(g^{-1}{\rm d}g)^{4n+1} ,
\nonumber
\end{alignat}
where $d=4n+2$.
Let us estimate the above in the case ${\rm M}={\rm S}^d$  
($\partial{\rm M}_+={\rm S}^{d-1}$) for simplicity.
Notice that generic U($2N$) gauge transformation $g$ can be decomposed into 
U(1)$\times$SU($2N$) such that $g(y)=e^{i\phi(y)}\tilde g(y)$ where $\det\tilde g=1$.
The time reversal constraint tells that $\phi(-y)=-\phi(y)$ mod $2\pi$.
In $d=2$ (i.e., $n=0$), $\Delta_1[g]$ is given by this U(1) part,
$\Delta_1[g]=-N/\pi\oint {\rm d}\phi$, where line integral is over ${\rm S}^1$, namely, 
the equator of ${\rm S}^2$. This gives manifestly an even integer.
On the other hand, in higher dimensions, contribution from U(1)
vanishes and only
the non-Abelian sector $\tilde g$ enters into $\Delta_{d-1}$;
$\Delta_{d-1}[g]=\Delta_{d-1}[\tilde g]$. 
The time reversal constraint tells that at the time reversal invariant 
points $y_j$ ($j=1,2$ on ${\rm S}^{d-1}$), $\tilde g(y_j)\equiv h(y_j)\in$ Sp($N$) for 
$d=8n+2$, whereas $h(y_j)\in$ O($2N$) \cite{footnote3} for $d=8n+6$.
First, let us consider the former case.
Assume that $\tilde g$ takes $\tilde g_0\notin$ Sp($2N$) at a certain $y$, 
$\tilde g(y)=\tilde g_0$. 
Then, $y$ cannot be $y_j$, and the time reversal constraint ensures that at 
$\tilde g(-y)=\tilde g_0$.
There are thus even number of points on ${\rm S}^{d-1}$ which are mapped to $\tilde g_0$. 
It turns out that the degree of the map $\tilde g$ is even, implying that 
the winding number, $\Delta_{d-1}[\tilde g]$, is even. 
On the other hand, 
if one cannot find any $\tilde g_0\notin$ Sp($N$) on ${\rm S}^{d-1}$, namely, 
if $\tilde g(y)\in$ Sp($N$) for all $y$, $\Delta_{d-1}[\tilde g]=0$. 
We thus conclude that $\Delta_{d-1}[g]$ is an even integer for $d=8n+2$.
The case $d=8n+6$ ($n=0,1,\cdots$) is likewise.

If the gauge potential can be smooth on the whole ${\rm M_+}$, $D_d$ should be zero,
which is a trivial element of Z$_2$. 
Now we shall show that there exist not only such a trivial element but also 
a nontrivial element indeed.  
We assume that the u($2N$) gauge potential is $2\times2$ block-diagonal.
Then, the time reversal invariance (\ref{TimRevInvGau}) requires that 
the upper and lower u($N$) sector of the 
gauge potential is not independent, given generically by the form
\begin{alignat}1
{\cal A}_\mu(y)=
\left(\begin{array}{cc}a_\mu(y)&\\&a_\mu^*(-y)\end{array}\right) ,
\label{GauPotDec}
\end{alignat}
where $a_\mu(y)$ denotes a u($N$) gauge potential.
In this block-diagonal case, since the upper and lower sectors are decoupled, 
the Z$_2$ index and $D_d$ can be separately computed such that
${\rm ind}_+i\slashchar{D}={\rm ind}_+i\slashchar{D}_\uparrow+{\rm ind}_+i\slashchar{D}_\downarrow$
and $D_d=D_{d\uparrow}+D_{d\downarrow}$, where arrows mean that only the upper ($\uparrow$) or
lower ($\downarrow$) gauge potential is taken into account.
Assume that the gauge potential in the upper sector is nontrivial,
which yields a nonzero index and Chern number. 
This means that $a_\mu(y)$ cannot be smooth over ${\rm S}^d$:
For simplicity, assume that we have two kinds of gauge such that 
$a_\mu^{(\pm)}(y)$ is regular in ${\rm M}_\pm$, collecting all singularities in $M_\mp$.
The ordinary index theorem \cite{footnote4} 
claims that ${\rm ind}\,i\slashchar{D}_\uparrow=N_+-N_-$,
where $N_\pm$ is the number of the zero-mode with the chirality $\pm$.
Now let us choose $a_\mu^{(-)}(y)$ as the upper gauge potential.
Then, Eq. (\ref{Z2IndThe}) gives $D_{d\uparrow}={\rm ind}\,i\slashchar{D}_\uparrow=N_+-N_-$, since
in this case, $a_\mu^{(-)}(y)$ is regular in ${\rm M}_-$ and therefore,
\begin{alignat}1
-\int_{\partial{\rm M}_+}\omega_{d-1}({\cal A})=\int_{\partial{\rm M}_-}\omega_{d-1}({\cal A})=
\int_{{\rm M}_-}{\rm tr}{\cal F}^{d/2} ,
\nonumber
\end{alignat} 
holds if the lower gauge potential is neglected.
Next, let us switch to the case with lower gauge potential only, 
which should be $a_\mu^{(-)*}(-y)$.
Since this gauge potential is regular in ${\rm M}_+$, 
it never contribute to Eq. (\ref{Z2IndThe});
$D_{d\downarrow}=0$. 
It thus turns out that $D_d=N_+-N_-$ holds for the full gauge potential ${\cal A}_\mu$ 
in Eq. (\ref{GauPotDec}). 
On the other hand, since $i\slashchar{D}_\downarrow$ has $N_\mp$ zero-mode 
with chirality $\pm$,
i.e., ${\rm ind}\,i\slashchar{D}_\downarrow=N_--N_+$ \cite{footnote4}, 
we reach ${\rm ind}_+\, i\slashchar{D}=N_++N_-$.
Therefore, we conclude that ${\rm ind}_+\,i\slashchar{D}=D_d$ mod 2.

Let us now take into account off-diagonal elements of the
gauge potential. The singularities in the upper gauge potential 
may move but stay in ${\rm M}_+$ 
if off-diagonal elements are small enough. 
Even if one of them moves into ${\rm M}_-$, its partner
in the lower gauge potential in ${\rm M}_-$ moves into ${\rm M}_+$. 
This is due to Eq. (\ref{TimRevInvGau}) which ensures that 
if ${\cal A}_\mu$ has a singularity at $y$, an opposite singularity
appear at $-y$. Therefore, $D_d$ can change only by 2.
On the other hand, along the change of the gauge potential, the spectrum of the 
Dirac operator flows, and a nonzero-mode quartet can be two zero-mode doublets
and vice versa, which result in the change of Z$_2$ index also by 2.
From the point of view of such moving singularities, the mod 2 gauge-dependence of 
$D_d$ can be understood likewise. It thus turns out that $D_d$ and the Z$_2$ index
change by two and therefore coincide mod 2.
 
Finally, we shall exemplify a Dirac operator with nontrivial Z$_2$ index in $d=2$.
Let us consider a Dirac operator on ${\rm S}^2$ with magnetic monopole background fields \cite{EGH80},
\begin{alignat}1
i\slashchar{D}(\theta,\phi)
=i\sigma^1\left(\partial_\theta+\frac{1}{2}\cot\theta-i{\cal A}_\theta\right)
+\frac{i\sigma^2}{\sin\theta}\left(\partial_\phi-i{\cal A}_\phi\right)
\nonumber
\end{alignat}
where $0\le\theta\le\pi$ and $-\pi\le\phi\le\pi$ are polar coordinates, 
${\cal A}_\theta$ and ${\cal A}_\phi$
are u(2) gauge potentials of the type (\ref{GauPotDec}), and the cotangent term is 
due to the spin connection \cite{EGH80}.
This Dirac operator has time reversal symmetry
\begin{alignat}1
{\cal T}i\slashchar{D}(\theta,\phi){\cal T}^{-1}=i\slashchar{D}(\pi-\theta,-\phi),
\nonumber
\end{alignat}
where ${\cal T}=\sigma^1i\tau^2{\cal K}$.
Two time reversal invariant points are
$(\theta,\phi)=(\pi/2,0)$ and $(\pi/2,\pi)$.
For the upper gauge potential we have two well-known possibilities,
\begin{alignat}1
a_\phi^{(\pm)}(\theta)=\frac{m}{2}(\pm1-\cos\theta),
\nonumber
\end{alignat}
and $a_\theta=0$, where $a_\phi^{(\pm)}$ is the charge-$m$ monopole potential 
with a singularity at the south and the north pole, respectively.
From Eq. (\ref{GauPotDec}) it follows that the lower potential should be
$a_\phi^{(\pm)*}(\pi-\theta)=-a_\phi^{(\mp)}(\theta)$,
telling that it denotes a monopole with the opposite charge and 
with the singularity at the opposite pole.

Assume $m\ge0$ and choose $a_\phi^{(-)}$ as the upper gauge potential.
Then, $i\slashchar{D}_\uparrow$ gives just $m$ zero-modes with chirality $+$,
whereas $i\slashchar{D}_\downarrow$ gives the same $m$ zero-modes but with chirality $-$.
On the other hand, we see $D_{d\uparrow}=m$ and $D_{d\downarrow}=0$.
Therefore, for this decoupled model and the present gauge-fixing, 
the Z$_2$ index and $D_2$ coincide, ${\rm ind}_+\,i\slashchar{D}=m=D_2$.
However, as discussed, these can change by 2 
by gauge transformations and/or deformation of the gauge potential,
and generically coincide modulo 2.

We would like to thank H. Oshima for fruitful discussions.
This work was supported in part by Grant-in-Aid for Scientific Research
(No. 20340098 and No. 21540378).


\begin{thebibliography}{99}
\bibitem{footnote1}
Even in conventional imaginary-time Euclidean space,
we can also define a Z$_2$ index of the Dirac operator similarly in the text:
Time reversal invariance is given by
${\cal T}\slashchar{D}(x){\cal T}^{-1}=-\slashchar{D}(x),$ where
$\slashchar{D}(x)$ is the hermitian Dirac operator and $x$ denotes the coordinates
of imaginary-time Euclidean space.
By the use of the anti-commutativity of $\slashchar{D}$ and $\Gamma_5$, we see that 
a similar discussion of the quartet formation at nonzero energies can apply.
However, in this case, it is not possible, as far as we study,
to find a corresponding topological invariant.
\bibitem{footnote2}
This choice of $J_1$ is just for a practical reason:
It allows us to give a nontrivial example of models by the use of 
Eq. (\ref{GauPotDec}).
As far as symmetry is concerned, we can choose $J_1=1$. See \cite{footnote3}.
\bibitem{footnote3}
In a suitable basis, $\tau^1$ becomes diagonal $\tau^3$. Then, 
we see $h\in$ O($N$,$N$,C) $\simeq$ O($2N$,C). It thus turns out 
O($2N$,C)$\cap$ SU($2N$)$=$O($2N$).
\end{thebibliography}
\end{document}